\documentclass[letterpaper]{article} 
\usepackage[]{aaai24}  
\usepackage{times}  
\usepackage{helvet}  
\usepackage{courier}  
\usepackage[hyphens]{url}  
\usepackage{graphicx} 
\usepackage{natbib}  
\usepackage{caption} 
\usepackage{algorithm}
\usepackage{algorithmic}
\usepackage{newfloat}
\usepackage{listings}
\DeclareCaptionStyle{ruled}{labelfont=normalfont,labelsep=colon,strut=off} 
\floatstyle{ruled}
\newfloat{listing}{tb}{lst}{}
\floatname{listing}{Listing}
\title{Strategies for Increasing Corporate Responsible AI Prioritization}
\author {
    Angelina Wang\textsuperscript{\rm 1},
    Teresa Datta\textsuperscript{\rm 2},
    John P. Dickerson\textsuperscript{\rm 2, 3}
}
\affiliations {
    \textsuperscript{\rm 1}Stanford University\\
    \textsuperscript{\rm 2}Arthur\\
    \textsuperscript{\rm 3}University of Maryland
}

\usepackage{bibentry}

\begin{document}

\maketitle

\begin{abstract}
Responsible artificial intelligence (RAI) is increasingly recognized as a critical concern. However, the level of corporate RAI prioritization has not kept pace. In this work, we conduct 16 semi-structured interviews with practitioners to investigate what has historically motivated companies to increase the prioritization of RAI. What emerges is a complex story of conflicting and varied factors, but we bring structure to the narrative by highlighting the different strategies available to employ, and point to the actors with access to each. While there are no guaranteed steps for increasing RAI prioritization, we paint the current landscape of motivators so that practitioners can learn from each other, and put forth our own selection of promising directions forward.
\end{abstract}

\section{Introduction}
There is increasingly research on how to implement fair and responsible artificial intelligence and algorithmic bias techniques. However, relatively little of this research has translated into industry practice~\cite{holstein2019industry}, and recent research has often found that a key obstacle is institutional barriers which hamper adoption~\cite{madaio2022assessing, madaio2020checklists}.
In an ideal world we might have both better-scoped and higher quantities of regulation in place, and while there are certainly movements in this direction (e.g., EU AI Act, GDPR), policymaking can be slow and imprecise. In the meantime it is important to understand the motivations that companies have historically had for incorporating responsible AI concerns so that we can learn from what has worked in the past in order to enact greater impact in the present.

\begin{figure*}
    \centering
    \includegraphics[width=0.96\textwidth]{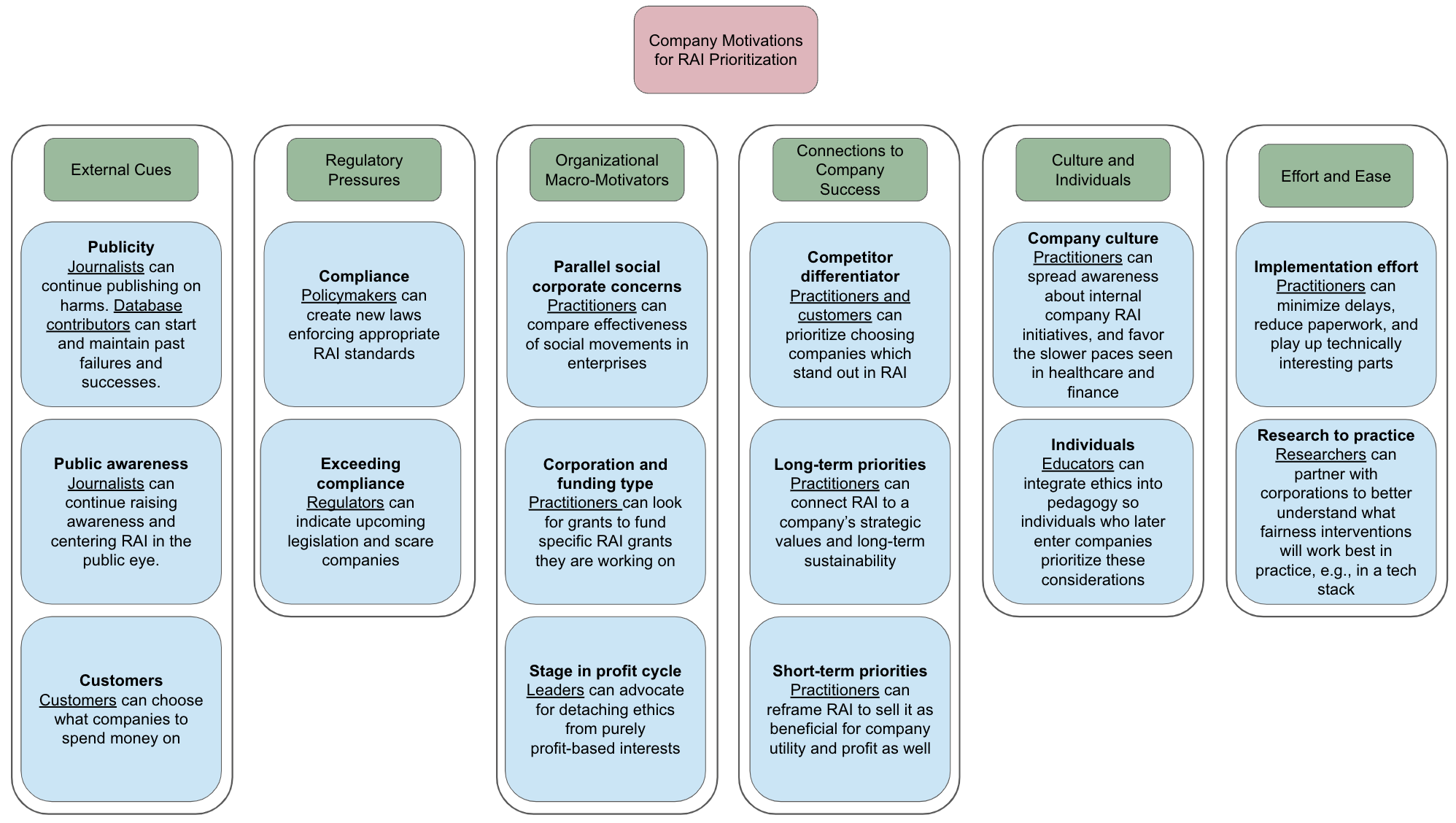}
    \caption{Six themes surfaced from our interviews as directions for increasing the motivation that companies have for prioritizing responsible AI. We summarize each of these six themes, and \underline{underline} examples of actors that are able to act on each.}
    \label{fig:summmary}
\end{figure*}

A major obstacle facing Responsible AI (RAI)\footnote{Throughout this work, we use the term RAI to describe the broad category of themes related to fair machine learning, AI ethics, and algorithmic bias. While it can be reductive to collapse all different notions of bias and fairness under the same terminology~\cite{blodgett2020nlpbias}, in our case we find this to be an appropriate level of abstraction, and will bring up specific instances where this abstraction hides relevant details.} in practice is the lack of corporate investment of resources. Individual RAI practitioners develop their own strategies within their respective companies~\cite{ali2023organizational}, but there is not yet a consolidated analysis of these separate techniques which could be shared with each other. The research question we seek to answer in this work is: \textit{How can we motivate companies to increase the resources they spend on RAI?} We foreground the word \textit{we}, and \underline{underline} throughout the work who has the power to pull on each of these levers that we unveil. To discover the strategies, we conduct 16 semi-structured interviews with participants who a) currently or in the past worked at companies that used algorithms with RAI-related concerns, and b) had insight into or control over the decision-making process that set RAI priorities. We provide a summary of our findings in Fig.~\ref{fig:summmary}, ultimately discovering six high-level reasons that motivate companies to prioritize RAI: \textit{external cues} like publicity, \textit{regulatory pressures}, \textit{organizational macro-motivators} like the funding type of a company, relevance to \textit{company success}, unique circumstances due to \textit{company culture and individuals}, and the \textit{effort and ease} of implementation. 
It boils down to prioritization: companies are rarely against RAI, but when there are many competing interests, RAI is often neglected.

However, the story is not as simple as ``increase regulation'' or ``publicize RAI failures,'' strategies which prior work has covered as well~\cite{ali2023organizational}. Rather, there is nuance to each of these strategies, and the \textit{details} are the strength of our work. We provide a novel perspective by revealing the complexity behind each strategy as well as the actors most able to exercise them. For example, while public RAI failures are known to motivate prioritization, they can also scare companies from engaging in RAI at all, fearing backlash for an attempt that doesn't meet a journalist's standards. Additionally, while we know that companies are scared of not complying with legislation for lawsuits, our interviews provide the novel insight that companies are not just scared of legitimate lawsuits, but rather any publicity that could lead to a frivolous lawsuit because of what might be surfaced during the legal discovery phase. Or while we might have an idea of how RAI can serve as a company differentiator, the critical impact of generative AI on this differentiation has not been previously examined. Our interviews unveil that with the rise of generative AI, differentiating via RAI provides an arena for companies that cannot afford to compete on compute-heavy tasks.

Ultimately, our interviews detail a tangled web of ambiguity: there are no guaranteed steps forward for increasing RAI prioritization (not even regulation, which can be heavy-handed, misdirected, and hard to get right). Yet, by painting the landscape of strategies and showing the challenges within each, we bring clarity and structure to encouraging ways forward. As an example, a practitioner who is having trouble getting buy-in for an RAI-related concern learn that they can leverage the argument that it will help their company differentiate from a competitor. Additionally, by naming actors with power to enact change, we offer a call to action to previously overlooked groups, such as venture capitalists and consultants. We conclude by discussing four concrete directions that emerge to us as the most promising.
By looking to the past we scope ourselves to the efforts that have already been tried. However, we believe that through this consolidation of separately-performed techniques, we can illuminate the landscape of approaches for RAI prioritization and help to inspire novel methods going forward.
As anonymized participant \#4 (P4) puts it, ``There's a lot of really great ways that people in the field can meet each other to navigate some of the challenges that we all share.''

\section{Related Work}

\subsection{AI Ethics in Industry}

There are many difficulties to implementing RAI in practice. 
These are due to factors like organizational barriers~\cite{deng2023crossfunctional, rakova2021organizational, madaio2020checklists}, principles which are hard to operationalize for reasons such as being too vague~\cite{holstein2019industry, schiff2020gap, schiff2021gap, mittelstadt2019principles, munn2023uselessness, green2020falsepromise}, research outputs which do not match practitioner needs~\cite{madaio2020checklists, madaio2022assessing}, and the structure of labor and workplaces~\cite{shilton2012valueslevers, widder2023dislocated}.
Of these obstacles, our work focuses on the lack of resources invested by companies.

Many important works have explored the dynamics between RAI and corporations under capitalism. \citet{ali2023organizational} conduct a thorough analysis of RAI implementation in technology companies. They find three key barriers: the overriding authority of software product launches, the difficulty of quantifying ethics, and the frequent team reorganization that can cause loss of knowledge. 
Rather than focusing on the barriers RAI practitioners have, in our work we examine RAI successes to understand what can be amplified to increase RAI buy-in. Both of our works share the goal of suggesting steps forward to increase RAI investment, but due to the different framing, settle on different ways forward that are collectively important. \citet{metcalf2019ethics} capture how the inherent culture of Silicon Valley can be adversarial to substantive ethics due to differing incentive structures and sometimes-conflicting values, and \citet{phan2022virtue} describe the conceptualization of ethics knowledge as simply another form of capital that is exchanged. On the other hand, recent work describes a shift in perspective that a company's responsibility encompasses social impact in addition to just profit---though potentially only larger companies have the resources to do this, compared to smaller ones~\cite{dearteaga2022business}. Other works look specifically to smaller startups and the unique pressures they face in this domain~\cite{winecoff2022startups, hopkins2021resource, vakkuri2020prototype}.

Ultimately while ways of increasing RAI prioritization have surfaced in prior work, our singular focus on this question brings out new details and nuances, and can help to inform \textit{who} can do \textit{what} in increasing RAI prioritization.

\subsection{Organizational Theory}
In our work, we focus on inductively discovered findings rather than deductive, theory-driven themes, but provide discussion of relevant broader theories here.
Central is the economic system of capitalism allowing private actors to operate with a central motivation of maximizing profit~\cite{gilpin2000challenge}. In theory, this motive for profit is said to ensure the efficient allocations of resources~\cite{smith2002inquiry}. However, this market fundamentalism can play out in many ways: scholarship has examined how organizational policies are often decoupled from daily practices~\cite{meyer1977institutionalized, bromley2012smoke}, and our work examines similar ongoing tensions like between valuing ethics in the abstract and prioritizing profits in practice. 
        

Neo-institutional theory is one of the key frameworks for understanding organizational behavior and change~\cite{powell2012new}, and posits that sources of organizational change can be endogenous or exogenous, i.e., internally or externally-triggered. 
Actors in an organization can endogenously construct change, and our work sheds lights on the specifics of this in the RAI space. Through the lens of organizational field theory, organizations can internally forge new paths through the interactions of the inhabitants of these institutions~\cite{schneiberg2007s, hallett2006inhabited}. Additionally, corporations can be reordered by specific highly-connected individuals who take strides to disrupt the old institutional structures~\cite{weber1978economy, managing_legitimacy, dimaggioagency}. 
Meanwhile, exogenous sources include environmental ``jolts'' (e.g., sudden societal events), force organizations to adapt in unique ways. These ``transitional blips'' allow for the strengthening of cultural priorities and introduction of unrelated changes~\cite{jolts}. We discuss a variety of external triggers for corporations: from publicity to social movements to regulations. Similarly, social movement literature discusses how corporate protests and large-scale media coverage affecting corporate stock prices can be particularly effective at inciting change~\cite{king2007social, davis1994social}. There are parallels to draw from domains like privacy and ESG (environmental, social impact, and corporate governance) issues to RAI's place within capitalism. In all of these spaces, a value besides profit comes into tension~\cite{hartzmark2023counterproductive, cohen2020esg}.


\section{Methods}
To answer our research question: \textit{Based on past successes, how can we motivate companies to prioritize RAI?}, we conducted semi-structured interviews with 16 participants from 13 companies.

\subsection{Participant Recruiting}
We sought participants who a) currently or in the past worked at companies that used algorithms with RAI-related concerns and b) had insight into or control over the decision-making process that set RAI priorities. We recruited these participants by reaching out to our personal networks, posting on Twitter, LinkedIn, Slack groups, and snowball sampling by asking participants to refer us to others who might be useful to talk to. In total, we interviewed 16 participants from 13 companies between July - September 2023, reaching thematic saturation by our last two interviews~\cite{green2004qualitative}. Given the specialized background of our participants, the pool of eligible participants was relatively small. Each participant was compensated with a \$25 gift card.
The industry and company size of each participant is listed in Tbl.~\ref{tab:participants}. Their roles are: scientist/engineer (10), product (2), manager/executive (4). Our participants span a wide range of domains: Software as a Service (SaaS), healthcare, finance, manufacturing, and more. 

\begin{table*}
    \centering
    \begin{minipage}{.33\linewidth}
      \centering
        \begin{tabular}{c|c|c}
            Ptcpt. & Industry & \# Empl.\\
        \hline
        
        P1 & Healthcare & 10-100 \\
        P2 & Entertainment & 1k-50k\\
        P3 & E-commerce & 100-1k\\
        P4 & Finance & 50k+\\
        P5 & Manufacturing & 50k+\\
        P6 & Entertainment & 100-1k\\
        \end{tabular}
    \end{minipage}%
    \begin{minipage}{.33\linewidth}
      \centering
        \begin{tabular}{c|c|c}
            Ptcpt. & Industry & \# Empl.\\
             \hline
            P7 & Hiring & 10-100\\
            P8 & Entertainment & 50k+\\
            P9 & Finance & 50k+\\
            P10 & Electronics & 50k+\\
            P11 & [redacted] & 50k+\\
            & & \\
        \end{tabular}
    \end{minipage}%
        \begin{minipage}{.33\linewidth}
      \centering
        \begin{tabular}{c|c|c}
            Ptcpt.  & Industry & \# Empl.\\
             \hline
            P12 & Social Media & 1k-50k\\
            P13 & Finance & 50k+\\
            P14 & SaaS & 10-100\\
            P15 & Electronics & 50k+\\
            P16 & SaaS & 10-100 \\
            & & \\
        \end{tabular}
    \end{minipage}%
    \caption{Each participant (Ptcpt.) with their company's industry and employee number (\# Empl.). The roles are (data) scientist/engineer: 10, product: 2, manager/executive: 4.}
    \label{tab:participants}
\end{table*}

\subsection{Interview}
The first author conducted all of the interviews, which were 30-60 minutes and conducted on an online video platform. Each interview began with explaining the purpose of our study and asking whether participants were comfortable with the level of anonymity we would provide. We then asked participants if we could record the video and automatically generate a transcript; if not, the interviewer took extensive notes. The interview instrument of guiding questions is included in the Appendix. Generally, they were customized for each interview and depended on the participant's role at the company. 
We sent all participants a draft before submission to ensure comfort with their level of anonymity given the sensitivity of the subject. 

\subsection{Analysis}
We conducted an inductive, data-driven thematic analysis to identify the key themes in our 16 interviews~\cite{braun2006thematic, thomas2006inductive}. The first author coded all of the interviews, and picked out the six most theme-dense interviews which the second author coded as well. Then, the union of all codes were consolidated, resulting in 967 unique codes. These codes were relatively detailed (e.g., 
``Trade-offs are never explicit numbers of value, but things like how long can a team work on this,'' ``Team started out as engineers using canonical notions of fairness from papers, then as evolved got product and legal involved and asking more substantive questions''). The authors iteratively clustered the codes through five rounds of thematic groupings until consensus was reached on six high-level themes (section headers) composed of 15 sub-themes (subsection headers). 

\section{Results}
We surface six high-level themes about what motivates companies to prioritize RAI. \textit{External cues} and \textit{regulatory pressures} are frequently cited as the most powerful reasons. Four more arise as important and may be overlooked: \textit{organizational macro-motivators} like how a company is funded; relevance to \textit{company success}, company \textit{culture and individuals}, and \textit{effort and ease} of implementation. 
Topics in each category are not mutually exclusive, but we group them to bring structure to the strategies we have for affecting the prioritization that companies place on RAI.
Overlaid on these themes are the different actors that currently have power~\cite{widder2023power} and access to each lever (e.g., \underline{journalists} are relevant for impacting \textit{external cues}, but so are \underline{consumers}).\footnote{We do not touch on collective action or activism~\cite{belfield2021activism} as none of our participants brought it up, but it is a critical component of power~\cite{widder2023power}.} We \underline{underline} the named actors throughout, and in Appendix Fig.~\ref{fig:stakeholder_summary} summarize these group-specific strategies by difficulty of implementation. Of course, these are not the only people able to pull these levers, just some of the closest.

Ultimately we find a complex web of factors, and acting on one or even a set of strategies is not guaranteed to increase RAI prioritization. 
However, while we are careful to warn against a confinement to these options which would foreclose additional strategies, our analysis of previously successful motivators helps to inspire a promising path forward and equip practitioners with a toolkit of strategies.

\subsection{External Cues}
One of the largest motivators voiced by participants to prioritize RAI was external cues. This refers to axes like media publicity, general public awareness, and customer demands.

\subsubsection{Publicity}
Publicity, and specifically the fear of bad press, was brought up by every participant as a major motivating factor. As P14 mentions, the ``impetus for change usually comes when something bad happens, right?'' 
P[4, 5, 8, 9, 11, 15] all specifically bring up the utility of having a ready list of public failures of other companies, and P15 explains, ``the most effective thing was pointing to failures of other companies.''
P4 echoes that they are ``a wealth of knowledge of cautionary tales that help me make those conversations and get that buy-in,'' since examples convey how even when you think you have control, things can always go wrong. The Partnership on AI's AI Incident Database\footnote{\url{https://partnershiponai.org/workstream/ai-incidents-database/}} was brought up by a few participants as being useful for this.


However, as great of a motivator as bad publicity is, there are also downsides. P12 mentions that even though their company was working on RAI, they were too scared to publicly talk about it or implement anything that would make those efforts visible, lest that lead to bad publicity. Their company preferred to remain silent on the topic rather than risk being in the public eye for claiming to be doing RAI work that could be criticized. Given how powerful of a motivator bad press is, it seems important to maintain it, but useful to bring nuance so that companies have the space to pursue RAI without facing quick and reductionist headlines.

In general, good press did not seem to be a very strong motivator. The only notable instance was a public ethics award given to one participant's organization, which they cited as a motivator for the company to live up to the award. 
Even though rewarding behavior that should be expected is not ideal, creating reward incentives for companies could help communicate it as something the public values. This is analogous to the LEED rating system or Fair Trade labels, though all have been shown to be susceptible to different forms of ethics-washing~\cite{bowen2014greenwash, doan2017green, low2005roast}.

Overall, \underline{journalists} have immense power in carefully exposing irresponsible AI practices at companies, and \underline{maintainers} and \underline{contributors} of incident databases give practitioners an arsenal of arguments to draw from. However, concerningly, there is a decline in journalism funding~\cite{bauder2022newspapers}.

\subsubsection{Public awareness}
Downstream of journalism is the growing public awareness that AI is not objective and can cause disparate harm. P[1, 3, 4, 5, 7, 8, 11, 12, 14, 15] all bring up how RAI has entered the public consciousness, and that this collective zeitgeist has facilitated getting company buy-in on RAI. P14 notes, however, that they believe RAI is becoming less trendy than it once was. In the same way that social justice movements (e.g., \#MeToo, \#BlackLivesMatter) can lose momentum, we need to collectively ensure these topics are not just a passing trend. Again, \underline{journalists} will play a critical role in maintaining relevance in public awareness.

This public awareness can also have other positive effects for RAI: P8 brings up the externality of ``even if those obvious problems don't have a legal hook yet, because the more things are in the press, the more likely someone will sue. Even if those lawsuits might not have great standing or grounds, just for the optics of it or if those lawsuits get to a discovery phase... it could lead to embarrassing things being uncovered.'' 
Public awareness can also affect the recommendations that \underline{consultants} and other influential external advisors provide. P4 shares anecdotally that they were told ``it's a lot easier as an external advisory function or a provider or a partner to be able to push... to get some things done versus internally.'' P10 also describes how external consultants provide recommendations based on what they project to be future trends, and for such a speculative enquiry, matters that are high on the public radar can become a self-fulfilling prophecy.
Many participants also brought up large language models and the increase in generative AI as leading to an increase in awareness around the capabilities and harms of these models (P[4-13, 15]). 
However, in many cases the excitement to develop and deploy these generative models outweighed the increase in concern. As P11 describes, ``It mostly pushed the innovation impulse... So it did increase the number of people who saw the salience [of RAI], but it didn't necessarily give it on balance more power because it got less momentum than the business.'' 

At this point, our conflation in wording by using ``RAI'' as a catch-all term runs into collisions. P5 brings this up by differentiating between those components of RAI that are more known and thus prioritized, compared to lesser-understood components which are left behind. They explain ``things like fairness, privacy, those are already very high awareness so we get lots of great traction around conducting privacy impact assessments and around doing bias testing and performance testing. But when it comes to things like accountability and transparency and labor impacts, that gets a little murkier.'' Thus, now that we have established the first level of public awareness, \underline{internal advocates} can help move towards more nuanced notions of RAI that encompass components like labor impact.


\subsubsection{Customers}
The third aspect of \textit{external cues} is the power of the customer\footnote{We use the term ``customer'' to include both consumers in customer-facing apps as well as clients in business-to-business organizations.} (P[1-3, 5-14]). This is relevant both for retaining current customers as well as attracting new ones. However, it did seem to be differently prioritized by companies across domains. For example, healthcare and finance are two domains that stand out as caring more about customer perception of RAI: 
P9 says
``as a bank, if you damage your brand, people will not trust you as much as they did before.'' 
Another relevant axis for how much customer demand can drive RAI is the type of customers desired. P7 notes that ``if you don't build trust among underserved communities, and among small companies who basically cannot afford to have bias in their systems, we will not make profits, we won't get customers.'' This is similar to how prior work has found that startups need to work harder than larger companies in establishing trust and legitimacy~\cite{winecoff2022startups}.

One way customers can express demand is through their purchasing behavior and technology usage. P8 mentions boycotts and grassroots campaigns that have shifted company prioritization, and P14 mentions successful examples of other companies in the marketplace.
In the technology space, companies built on the customer demand of privacy include DuckDuckGo and Signal. \underline{Customers} can exercise signaling behaviors with their capital in order to communicate to companies their desire for RAI. 
This serves as motivation to companies because, as P4 puts it, ``Why are you leaving customers on the table? Because of your unfair practices, right? You could catch more business.'' 

\subsection{Regulatory Pressures}
Beyond \textit{external cues}, the other most frequently cited motivator for prioritization of RAI were \textit{regulatory pressures}. 

\subsubsection{Compliance}
\label{sec:follow_law}
First and foremost, nearly all of our participants bring up the relevance of legal compliance as a reason to prioritize RAI (P[0-5, 7-15]). 
More specifically, it is the fine:
``If you do something and your company is liable, it's billions [of dollars], so actually you need to take care of these things'' (P10). 

However, P[1, 8, 11, 12, 15] also bring up situations where the presence of regulatory pressures, whether real or imagined, actually hamper RAI adoption. One such well-studied obstacle is that of regulation prohibiting the collection and/or usage of sensitive attributes, which is often necessary in RAI work to even measure the amount of unfairness that is present~\cite{ho2020affirmative, kumar2022credit, bogen2020awareness, andrus2021demographic}. 
P15 mentions that they find legal work-arounds, for example collecting pronouns instead of gender. 
P12 also mentions that in many cases it is very unclear what the regulation is asking for and prohibiting, and so trying to adhere to it is itself a challenge. 

Ultimately there is a somewhat mixed bag with respect to whether more regulation is desired or not.
P8 notes two reasons more regulation might actually be harmful to RAI goals: ``uncertainty over whether the act of engaging in the work itself might be creating risk''
and ``if something is required in all cases in a blanket manner where that requirement is actually not well-suited for a certain circumstance, but people have to do it anyway, that leads to a lot of confusion.'' Echoing the second point, P12 adds that in lower risk domains, more legal regulation could be harmful by adding extra overhead with small impact. \underline{Policymakers} can spend more time with experts to understand the specific domains where more regulation can help with RAI prioritization, and how to scope it.

\subsubsection{Exceeding Compliance}
In addition to merely adhering to regulation, we find two reasons that regulatory pressures may motivate companies to do more than the law explicitly asks.
The first is to prepare for future regulation that may be more strict. For example, P13 points out that 
``we think `hey, there might be change in the future,' so we already have systems in place where we already are following a much stricter regulation that the one that already exists.'' Conversely, P4 mentions that because regulation might change in ways you cannot anticipate, there is a negative incentive to go above and beyond in case what you do conflicts with future regulation. The second reason is to demonstrate to regulators that self-regulation is sufficient, and to fend off stricter impositions. By demonstrating responsibility, companies hope to have a seat at the table when regulation is discussed. For example, P8 says ``failure of private actors to act of their own volition is likely to lead to more strict requirements... those requirements might end up being more difficult to achieve than what the voluntary work would have done.''

There are a number of ways to interpret these statements, and one takeaway is that even signaling done by \underline{regulators} about future regulation can help motivate action. However, given the ambiguity that regulation can take, it is important to communicate it in ways that encourage benevolent action, and are accompanied by substantive follow-through. 

\subsection{Organizational Macro-Motivators}
\label{sec:broader_structures}
Participants also described broader organizational factors for pursuing RAI which range from taking advantage of parallel social movements like environmental sustainability to external market forces and company profitability. 

\subsubsection{Parallel social corporate concerns}

Participants cite being able to leverage growing enthusiasm around other social movements, namely environmental sustainability and data privacy to boost RAI. P[4, 5, 8, 10] describe the benefits of leveraging the paths paved by environmental movements in industry, such as carbon footprint reporting, environmental impact assessments, and ESG investing. These structures have helped define the role of expertise for environmental activists, providing them with not only a voice but also authority to weigh in on trade-offs and decisions. P4 described leveraging executive commitments to environmental efforts, ``If you can connect it to those sustainability conversations, I think that has the most traction right now.'' 
\underline{Practitioners} can understand what tactics worked best in parallel spaces to better define the roles and authority of expertise for RAI.


\subsubsection{Corporation and funding type}
Structural financial aspects of an organization also affect the ability to implement RAI practices. Both P7 and P14 cite non-profit compared to for-profit organizations as more conducive places for prioritizing RAI principles. One participant's company is registered as a Public Benefit Corporation, indicating that their company mandate is to specifically benefit the public in some
way.\footnote{Public Benefit Corporations are different from a ``B Corp,'' or a Certified B Corporation, which is a third-party certification of social and environmental standards operated by the non-profit B Lab~\cite{bcorp_koehn}.}  Outside of the corporation's filing status, P1 and P7 also discuss the additional flexibility that receiving funding from external grants allows. For these participants, grants were used as a funding source for certain RAI-related projects which allowed for longer explorations into aspects of fairness and equity. 
\underline{Founders} can exercise choice in signaling their mission through their corporation type and filing status, and \underline{RAI advocates} can look for grants to fund specific projects they are working on.

\subsubsection{Stage in profit cycle}

There are differing financial pressures depending on the maturity of an organization. P9 and P15 both described the benefits of larger, more established companies having the space to focus on maintenance and impact in a way that smaller companies do not. P1 also describes, ``it really depends on where the company and product is, and it's lifecycle. [That helps decide] how much you invest [on RAI].'' Meanwhile, for startups, their stage in the company life cycle matters. Startups that are not currently raising capital may have more room to pursue principles of impact, as P7 describes, they ``don't have to appease investors or expectations of profit'' as heavily as they might have had to when seeking to raise a new round of funds.

Relatedly, P[5, 7, 12] specifically discussed recent downward economic forces resulting in company-wide hiring pauses and ``an organizational environment in which people just have more on their plate and it's easier for things to fall through the cracks because there is incentive to focus on different things'' (P5). This makes it difficult for RAI efforts to grow or even continue. Conversely, when the market is good, P[1, 2, 5, 10, 12] all cited greater access to resources ``to focus on impact rather than revenue'' (P1). RAI efforts often rise when the economy is good and ebb when the economy is bad. \underline{Venture capitalists} can be variance reducers here, supporting RAI in times of both wealth and austerity.

\subsection{Company Success}
Individuals frequently spend effort connecting RAI to company success to increase buy-in within the scope of already-existing corporate incentives~
\cite{ali2023organizational, metcalf2019ethics, phan2022virtue, rakova2021organizational}. P[4, 5, 8, 15] brought up looking at their company history to see what worked best in the past (e.g., ``finding examples of where we've seen positive changes in our operations, and really do audits of how did they get there?'' - P4). While some strategies will be company-specific, part of the motivation of this work is to consolidate narratives so RAI practitioners can learn from each other. All of the strategies in this section are ways for \underline{internal practitioners} to frame RAI for increased prioritization.

\subsubsection{Competitor Differentiator}
One powerful motivator for companies to prioritize RAI is as a differentiator from competitors~\cite{widder2023dislocated}. Many participants specifically brought up publicity around companies that have mistreated or gotten rid of their RAI teams as a motivator for companies to set themselves apart (P[2, 4, 5, 7, 10, 12, 15]). P5 describes that their company is trying to ``establish [themselves] in a position of thought leadership. But also distinguish ourselves against the types of FAANG companies that have made very visible pivots away from investing in these kinds of efforts.''
P10 adds that differentiating based on RAI offerings rather than aspects like generative AI can be more appealing because it is easier to compete with larger, more well-sourced companies on this front.


However, the incentive to use RAI as a peer differentiator is not guaranteed to lead to more responsible AI. As P8 comments, ``all the companies are now trying to compete on demonstrating responsibility for more advanced forms of AI. But that's very undefined, and so, you don't have to do much to try to demonstrate. There's no one who can call you out on saying that's not enough because no one agrees on what's enough.'' This indicates an urgent need for \underline{journalists} to partner with \underline{experts} to weigh in for the public on which company's RAI efforts are substantive.
We see the incentive to ethics-wash throughout our work, and will elaborate on this more during our recommendations.

\subsubsection{Long-term priorities}
Another route practitioners take to tie RAI into company success is through alignment with long-term priorities. For example, through saving time down the line by incorporating RAI now.
P8 motivates that ``scrambling to do something later on would be much more difficult than proactively adjusting course to do something that would be reasonable, and probably closer to the practices that the company was already doing.''
P10 invokes the idea of ``ethical debt''~\cite{cunningham1992techdebt, petrozzino2021ethicaldebt}, where even though incorporating ethics may take more time now, you will save time in the long run.

Another way is linking RAI to a company's strategic goals or values. P11 gives the example that ``I think the concept and the term is popular, and I think people wanted to be seen as engaging with it often for pro-social reasons... So if the CEO sends out a letter that says... we believe in rigor and quality. And then you say, this aligns with our company's goals of rigor and quality.''
P5 reports similarly that it was ``usually about reiterating some form of some value everybody has or that everybody should have or that everyone hopefully has, and then giving them recommendations that can help them be more in line with that goal.'' P6 points out how a specific company-level change helped with this, ``we went through and re-did our values at a company level, which is part of the reason why I think that I'm able now to talk about it because user-centricity is now one of our values when it wasn't explicit before.''


\subsubsection{Short-term priorities}
On the other hand, when a company decides how to prioritize large numbers of demands, short-term priorities can often beat out long-term ones, and these tend to, but do not always, deprioritize fairness.
P[1-3, 5, 6, 8, 10, 12, 14, 15] all bring up that ultimately RAI is secondary to utility of the product and financial success. P6 straightforwardly explains ``you need revenue to pay for the people who are actually going to solve the [RAI] problem,'' and
P13 laments ``I don't think any company at all is going to care that much about fairness or responsible AI until it starts hurting their wallets.'' For the most part, however, it is not that people do not care about RAI, but rather the relative urgency of RAI compared to other priorities feels smaller. 
The themes we discuss can help to create this urgency.

Thus, it often takes a shift in perspective or reframing of RAI under a utility perspective to elevate its priority (P[0-11, 13-15]). 
P5 gives an example where they told a team: 
``you guys who said that you need to collect as much data as possible... If we make things more explainable and transparent that will get people to talk more and that means more data for us.''
P4 argues for a bigger shift, and that when a perceived tension between RAI and utility comes up, ``why is this a compromise? Again, I think it's a failure of imagination... Are we going to hurt people in the process? I think there needs to be a shift in how we think about value.'' 
In terms of how personalized these reframings need to be, 
P11 describes that there was a ``standard battery of these concepts that we would draw on, but when you're talking to someone in model risk versus a data scientist, their needs and goals are very different,'' and P8 echoes there are ``people who truly deeply care about AI safety and could be convinced to care about fairness, but it's not through a lens of compliance. It's through a lens of AI safety and long term benefit to people.''
Even though this reframing needs to be catered per role and personal priority, techniques for these roles and value sets can be shared amongst \underline{RAI practitioners}.


One concrete strategy to ``sell RAI'' that practitioners found successful is tying together both numbers and a story (P[6, 11, 14]). As P6 explains, ``a truly complete story has both the numbers element and the emotional, look this is what a user actually experiences.''
Whereas our participants often find the story easier to craft, many express a need for better measurement techniques to formulate the number (P[4, 6, 8, 11, 12, 13, 14]). Fairness is notoriously hard to measure~\cite{jacobs2021measurement, wang2022representational, katzman2023taxonomy}, but there is an urgent need for interpretable, quantitative measures, both for decision-makers as well as customers (``a lot of the fairness issues are super technical and may or may not relate to the ability to explain to consumers whether they're being treated fairly or not.'' - P8). RAI measurements are important in communicating and explaining progress to non-technical stakeholders, who are often the decision-makers choosing what to prioritize.
P8 describes companies
``not really preferring to dive into really ambiguous spaces that could both alleviate and create risk without clarity.'' 
And yet, given that the ambiguity around RAI interventions will likely never go away entirely, we need to educate decision-makers about this inherent ambiguity and convey that it does not detract from RAI importance but does require more patience. As with many cases where it is contextual who is well-positioned to act on this, it could be an \underline{external consultant}, \underline{executive}, or \underline{coworker} who is best situated to communicate this. 

\subsection{Culture and Individuals}
Despite the common themes we've drawn from our interviews, our findings also reveal how subjective and context-specific the reasons for prioritizing RAI can be. 
Here, we discuss the role of company culture and individuals.

\subsubsection{Company culture}
\label{sec:company_culture}
Starting at the very high-level, country-specific cultural norms are one component that influence RAI prioritization. P10 works for a company based outside of the USA and mentions ``People don't change jobs in [country]... So practically their personality does get connected [with the company].'' They explain this means that an individual's sense of morality becomes intertwined with that of the company's, and they thus prioritize ethics.

Industry domain and company size are also relevant to company culture. Similar to how customers tend to be more valued in domains like healthcare and finance, P13 from finance notes that ``we're not like the other tech companies where we are constantly trying to be at the cutting edge'' and explains that this slowness allows better RAI integration. In terms of company size, P9 shares that compared to startups that may try to hire people who already have expertise, their larger company hopes to retain employees for a long time, and can afford to offer training sessions to teach already existing employees about RAI topics. 

Overall, many participants indicate that company culture feels ingrained and inherent, indicating it may be hard to change. In fact, P[1, 7, 10, 14] all specifically used the phrase ``DNA'' to describe company culture around RAI. Due to the often separate RAI teams and complicated organizational structures, participants also brought up the need to spread internal awareness for within-company RAI initiatives (P[5, 9-15]). Awareness of within-company RAI initiatives can help shift the culture, but often involves extra labor on the part of \underline{internal practitioners} to spread the word.


\subsubsection{Individuals}
\label{sec:individuals}
We find that often it is specific circumstances and individual experiences which cause people to champion RAI. We separate the impact of executives in more managerial roles compared to individual contributors.

Participants share that individuals higher up in the organization have some, but not full, power in increasing RAI prioritization. 
The cases where a single individual's desire has been enough is when they have sufficient power (P10), the company is hierarchical (P10), or the company is small enough and leadership is unified in their mission (P7). The circumstances leading to these individual efforts can be highly specific. For example, one participant mentioned their CEO attended the FAccT conference and was inspired. P16 also mentioned the relevance of personal risk: ``for executives, because it's not only company risk or losing your job, but it's personal brand risk.''
The other side of how personal these choices are is that individual circumstances can also cause RAI deprioritization. P8 points out two possible reasons: executives don't want to set a precedent for themselves that they will be held to in the future, and if previous RAI attempts did not pay off they may now have resistance. 
The unique circumstances that bring powerful individuals to RAI suggest it may be hard to implement any uniform strategy, but that it is still critical to create as many of these ``serendipitous'' situations as possible, e.g., by engaging more stakeholders at \underline{conferences} in the Responsible AI space, e.g., FAccT, AIES. One key intervention point is \underline{education}, as that is a formative time where many will encounter a set of pedagogy.

RAI work frequently falls to people of marginalized identities (P[1, 2, 5-7, 11, 16]) doing what is often volunteer work that does not help career advancement~\cite{deng2023crossfunctional, ali2023organizational}. 
P[4, 7, 11] mention the prevalent burnout amongst people working in this space, and P11 notes a unique challenge with respect to human resource problems. They explain that the same interpersonal problems are present in RAI career ladders as in other engineering career ladders, but unlike more generalized careers where someone can switch teams, there is often only one RAI team within a company, and thus no way to work under different management. This can lead to high turnover and losses in institutional knowledge (P5)~\cite{ali2023organizational}.

Another part of the problem is no one thinks RAI is their responsibility~\cite{widder2023dislocated, lancaster2023accountability}. P5 explains that ``engineers are very much in line with legal going `we just make stuff, I'm not responsible for what people do.' '' Individuals seem to believe that letting the default persist is not making a choice, when it is in fact a value-laden choice of its own.

As with many kinds of collective efforts, much of the success of RAI efforts boils down to interpersonal dynamics. P5 observes that ``there are some individuals internally... who have spent a lot of social capital to make [RAI] happen'' and P8 adds that there is a ``lot of internal maneuvering to get things through that are not built into a natural incentive structure in any organization.'' P11 elaborates that an important relevant factor is general pleasantness of working with someone even in terms of following up on emails.
They elaborate that 
``you can demonstrate being thoughtful, having good reasons, being a good collaborator over a period of time. Then even more resistant people, they'd sort of warm up... So, you're building out that your credibility, your reputation.''
P12 similarly mentions that their team focuses on being proactive with RAI metrics ready so any team that comes to them will not have to wait long.
It is unfair that RAI workers are held to a higher standard, which can contribute to the prevalent burnout. P[5, 11, 13] mention how RAI teams are overburdened and under-resourced, leading to additional strain.

Related to company ``DNA,'' individuals also sometimes think of being ethical as personally intrinsic. When asked about RAI adoption, many simply bring up the relevance of individuals being ``good'' (P[1, 6-12, 14]), and doing this work because they think it is the right thing to do or want to be seen as ``pro-social.'' P11 finds that in convincing people to adopt RAI, appealing to their morals often worked.
However, intention without expertise can also have downsides, and believing that inherent ``goodness'' is sufficient for RAI can be a serious harm. P8 flags that 
``sometimes people just start [RAI efforts] because they're interested and that can actually be problematic if they measure something that is using a definition a company's already decided is not the preferred one or has detriments.'' Thus, we need better \underline{education} that emphasizes how difficult RAI work is, and that it requires collaborating with experts. 

\subsection{Effort and Ease}
Finally, we consider how the effort and ease of RAI implementation in practice contribute to adoption likelihood. 

\subsubsection{Implementation effort}
Though perhaps obvious, many participants explicitly bring up how easier RAI efforts are far more likely to be adopted (P[4-6, 8, 10, 11, 13-15]). 
However, navigating ethical dilemmas is rarely easy, and necessarily requires time to both contemplate and implement. In practice, this often requires RAI teams to ``direct towards the more ethical path of least resistance'' (P4). P5 says that compromises are necessary in order to not be ignored: ``We're not in the business of saying `stop.' We try not to be a gate. We try to be a steering wheel.''
These responses can be demoralizing and feel like accepting the status quo rather than systemic change, and
P5 laments that ``you're never going to reach the radical thing that you want to reach,'' but explains this to be better than the counterfactual. There is a difficult balance in implementing (even incremental) improvements, even if they do not always solve the source problem~\cite{green2021dspolitical}. Sometimes though, speed and ease of an RAI intervention can be more of a rhetorical tool, and emphasized in framing rather than actual practice, by comparing it to what would take longer. P15 gives an example from trying to create an internal review system and ``trying to frame it as it's such a quick and easy process. All you have to do is [...]. [It] doesn't take that long and we really pared it down compared to what the traditional university does.''
Some other ways to make RAI intervention easier include tying into already existing pipelines. P[1, 2, 4, 10, 13] all voiced how it was beneficial to piggyback fairness issues onto other internal changes~\cite{deng2023crossfunctional}, e.g., P1 remembers ``taking advantage of the fact that we're integrating to new hardware as an opportunity to make this [RAI] change'' and P13 mentions ``if you're already spending a bunch of money to try and curate the data... then it makes sense to have the same team look at the data and make sure that data is also fair.''


We also find three key components of implementation difficulty that can influence individual support or resistance. Individuals do not like delays to deployment, detest paperwork, and do not find RAI interesting to work on. For delays in deployment, by the time a team encounters RAI concerns they often already feel their product is ready to be deployed, and see any delays as burdensome.
However, the seemingly simple solution of incorporating RAI from the start is hard in practice. Beyond just being difficult to change the status quo (P4), P13 paints the landscape: ``at any given point of time, there's just like thousands of teams working on thousands of different models. So it's basically impossible for [RAI folks] to be involved in the process of model creation right from the get-go. Because 80\% of the models that we work on don't ever see the light of day because they're all proof of concepts''~\cite{vakkuri2020prototype}.
On the other hand, delaying can also be a compromise. P12 describes using ``time-boxes'' where they convince teams to delay deployment by some predetermined number of weeks during which the team must demonstrate effort on implementing RAI interventions. They find this compromise useful, because teams have an end date assurance. However, it can be hard to enforce such effort-based approaches, and after deployment, effort is not a sufficient metric to end-users who suffer the consequences. 
Another key point of tension is paperwork. 
P5 shares that they use paperwork as both a carrot and a stick, offering to reduce the bureaucratic compliance paperwork for teams that can demonstrate they are incorporating RAI concerns: ``You can just kind of lean on people and annoy people with paperwork.''
Finally, P5 notes that some engineers just want to work on what they find technically interesting, to the point of sometimes even prioritizing that over what customers explicitly say that they want. 
P[1, 5, 6, 9-11, 13] all say that people see RAI as grunt work, and if it were intellectually interesting, it might be more prioritized. RAI presents genuinely challenging technical problems, so often this just needs to be made clear to practitioners. Through both framing and changes, if RAI interventions were presented by \underline{practitioners} in ways that seem to include justifiable and reasonable delays, do not increase seemingly useless paperwork, and are technically interesting, much of this individual resistance, which can be significant, would go away. And much of this can be in the framing---practitioners often do not feel paperwork is actually effective, so if it were streamlined to be more efficient and those filling it out were genuinely convinced about the importance (P11), these obstacles would be reduced.

\subsubsection{Research to practice}
\label{sec:r_to_p}
In translating RAI research into practice, a number of challenges are brought up by practitioners that impede RAI prioritization.
P12 discusses that academic research often doesn't consider things like the significant barriers of the engineering tech stack that make in-processing techniques far harder to implement compared to post-processing techniques, as the latter are easier to scale since they don't require coordination with large numbers of teams. 
In terms of the more principles-based approaches, P8 finds it's ``easy for the arguments to get theoretical and philosophical,'' and P4 says ``that's where I think the rubber meets the road, how do we actually get things done?''
Overall, prior work covers the numerous research-to-practice gaps in RAI~\cite{holstein2019industry}, and we bring this up as a \underline{research area} that can ease the adoption of RAI, and thus the willingness of companies to implement.

\section{Recommendations Forward}
\label{sec:future_directions}
Overall, we learn from our interview studies a wide range of ways to motivate companies to increase RAI prioritization. We present a summary of shared strategies in the Appendix (Fig.~\ref{fig:stakeholder_summary}), organized by the actor who is most able to enact each. From these, a few directions stand out to us as the more promising ways forward. Whereas we have endeavoured to remain mostly descriptive thus far, this section will be more prescriptive. 


\textbf{Accountability for ethics-washing.} A recurring theme is that in many cases, ethics-washing is more incentivized than substantive ethics work. To shift the incentives away from the former to the latter, we suggest greater collaboration between \underline{RAI experts and journalists} to hold companies accountable and make sense of cheaply made ethics statements. P8 and P10 both remark that performativity can actually lead to substantive action, e.g., ``The performative-ness of it actually does open space for more performance of it'' (P8). This is only true if companies are not rewarded for empty ethics statements, but also not unduly punished for publicizing genuine yet unsuccessful efforts.
Another way to communicate substantive RAI progress is by creating measurements which are more resistant to ethics-washing. These include measurements that incorporate human feedback and connect to more concrete harms~\cite{blodgett2020nlpbias,wang2022representational, wang2023stereotypes}. 

\textbf{Ethics modules.} Incorporating RAI requires two things on the part of an individual: intention and expertise. Both can be be taught through the integration of ethics modules during the education process that instill this as a priority, which may conflict with market pressures, early on~\cite{smith2023ethicseducation, fiesler2020techethics, raji2021pedagogy, grosz2019embeddedethics}. 
However, ethics modules in their current state 
may be overly focused on educating future executives and engineers on how they might identify and resolve ethics problems. Instead, the focus should be on communicating the difficulty and ongoing nature of ethics so that students have a genuine understanding and appreciation of the difficulty of the problem, and know how to collaborate with ethics experts rather than feeling falsely empowered to resolve these problems on their own. Like our finding that people sometimes feel that being a ``good person'' is a sufficient qualification to do RAI work, it's important to communicate that RAI and ethics work is a field that requires collaborating with those with expertise~\cite{mclennan2022embeddedethics}.

\textbf{Shifting structural incentives. }
We have seen alternatives to more traditional corporation structures such as nonprofits and public benefits corporations (PBCs), as well as indicators like ESG ratings.
In addition, there are new initiatives to support founders and investors prioritizing RAI.\footnote{\url{https://www.rilabs.org/}}
These have the potential to allow a company to be judged by shareholders for more than just their immediate profits, creating room for values like product quality and public perception~\cite{dorff2020pbc}.
There is a parallel to value-based healthcare~\cite{porter2006valuebased, teisberg2020valuebased}, which is a framework for healthcare that shifts the incentive structure to focus on patient health outcomes rather than cost reduction.
Of course, any of these could be used for ethics-washing, and prior work has postulated on why the supposed benefits of new forms of incorporation---signaling brand quality, limiting legal liability, and attracting capital---may not always materialize~\cite{bcorp_koehn}. However, these kinds of broader structural shifts can 
create an incentive structure that more highly prioritizes issues like RAI. They would require making the requirements for such legal qualifications and ratings strict enough to filter out disingenuous efforts, as well as potentially incorporating more financial benefits like tax incentives into PBCs and ESG ratings, instead of relying only on improved public perception.

\textbf{Research informed by needs. }
Participants shared that RAI research outputs often don't match their needs~\cite{holstein2019industry}. Two specific ways to close this gap emerge from our findings. One is that researchers can pursue integrated interventions that fit well into the typical software engineering stack. While this differs across companies, there are shared characteristics which have consistently made post-hoc interventions easier, and thus more likely, to be implemented (P12). This is in comparison to other interventions that require touching many parts of the tech stack. New research can pursue whether there are methods besides post-hoc that are also easier to fit into the stack. At the same time, this will require companies being more open about their internal processes and constraints. Another direction is in the interdisciplinary work happening between technology and the law. As participants expressed, it is often unclear which interventions are permitted by the law. P12 expressed that they were able to have more productive conversations with company lawyers after engaging with research that clarified the connection between law and concrete RAI approaches~\cite{ho2020affirmative, xiang2020reconciling, kumar2022credit}. Companies should also share ways of implementing RAI that is legally permissible with each other. For example, P15 mentions sometimes using pronouns instead of gender in RAI implementation, because while the latter is legally impermissible, the former is allowed. In this way, practitioners can share effort in navigating the complexity of regulation while pursuing substantive RAI efforts.


\section{Discussion and Limitations}
\label{sec:dis_limit}
Overall, in our work we unveil a set of strategies for affecting RAI prioritization. A limitation is that the recommendations may err on the side of being too accommodating to the world as it is rather than speaking to an ideal world---we do not wholly endorse all strategies so much as provide a slate of options from which to choose from. For example, is it right to say we should make fairness problems more technically interesting and ``fun'' in order to incentivize technical developers to work on them? Perhaps not, but in our endeavor of presenting all discovered methods of increasing RAI prioritization, it is certainly one possibility. As Audre Lorde famously put it ``For the master's tools will never dismantle the master's house. They may allow us temporarily to beat him at his own game, but they will never enable us to bring about genuine change''~\cite{lorde1984masterstools}. Ultimately by relying on the strategies that have historically worked within the confines of capitalism, our suggestions may be more of a patch than a treatment for the source problem. However, given the time it can take for more radical changes to take place, there is still benefit to working on increasing RAI prioritization in companies today, so long as it does not detract from such greater changes. RAI practitioners individually struggle towards this in their separate companies, and our work consolidates their techniques and efforts. Even though it seems like RAI will not be prioritized over profit, in our work we unveil that this is not clear-cut! There are successful strategies for stakeholders to elevate RAI even then, and the details of our work can be used as a toolkit for enacting this RAI prioritization. In the meantime, we should not lose hope for more radical changes and continue to push on that front as well~\cite{hampton2021oppression, davis2021reparation, green2020falsepromise, gorz1964labor}.


\section{Conclusion}

In this work, we look to the corporate landscape for examples of what has worked in prioritizing RAI in order to unveil the strategies at our disposal and pressure points to lean on in order to better motivate corporations under the landscape of capitalism. We find a complex picture of numerous actors able to exert pressure in numerous points, with no guaranteed outcomes. And yet, by drawing lessons from these past successes and applying enough force, we can collectively increase RAI prioritization.
However, if we believe that the current state of RAI adoption and prioritization is not sufficient, 
then it is clear that just relying on these strategies are not enough. While we believe this will be a useful toolkit for those looking to increase the prioritization of RAI, these strategies we unveil are just one set of the total possible interventions we should pursue. 

\section*{Research Ethics and Social Impact}

\textbf{Ethical considerations.} In this work, the biggest ethical consideration was the treatment of our participants and ensuring their comfort with their level of anonymity. This can be a sensitive topic that bears on someone's workplace environment, and we worked hard to ensure each participant felt comfortable with the level of anonymity they would be provided. To achieve this, we sought consent before either recording the interview or taking written notes. We also started each interview by asking each participant how they would prefer their background to be identified (e.g., job role, company industry, company size). Upon writing our draft and in response to some participants' concerns, we did not associate the role with each participant, as we found that our results were not specific to each person's role. Finally, we also sent the completed draft to each participant to ensure that they were ultimately comfortable with the level of anonymity they were represented with, since we understand that this can be a loaded space. We compensated each participant \$25 for an interview which ranged from 30-60 minutes.

All of the data was stored on Google Drive, which is secure and encrypted. We will delete all interview records upon final publication.

\textbf{Positionality.} One of the most relevant aspects of positionality for this piece is that two of the authors work in industry for a company in the RAI space, and one author is in academia. These positions affect the perspectives on corporations. In positions of relative socioeconomic privilege, none of the authors are personally susceptible to some of the more insidious effects of AI and so may find it easier to write about incremental, non-revolutionary improvements, compared to others.

\textbf{Adverse impact.} As discussed in Sec.~\ref{sec:dis_limit}, by having scoped our work to past efforts, we hope to not foreclose considerations of more radical change. However, we realize that papers of this sort that propose changes within existing capitalist structures could be seen as legitimizing the current structure, which is not our goal. We hope that our work can shine a light on directions forward, in parallel with other movements like greater instances of collective action or abolition in the cases that call for it.

\section{Acknowledgments}
We thank Namrata Mukhija and Daniel Nissani for helping with participant recruitment, Victoria Vassileva for initial conversations on these topics, and Amy Winecoff for feedback on the draft. This material is based upon work supported by the National Science Foundation Graduate Research Fellowship to AW, and was work initiated during AW's internship at Arthur.

\bibliography{references}

\appendix
\section{Interview Guide}

We conducted semi-structured interviews, and loosely started from the following questions. Most questions, however, were follow-ups to the responses given by the participants.
\newline

\noindent Background
\begin{itemize}
    \item What does your company do?
    \item What industry is your company in?
    \item What is the company size?
    \item What is your role?
    \item What is the level of Responsible AI practiced at the company?
\end{itemize}

\noindent Main Study
\begin{itemize}
    \item What are some of the primary motivating factors that cause RAI to be a priority?
    \item If there are organizational structures in place for responsible AI, e.g., an ethics review process before deployment, what do you think motivated this to be brought about?
    \item Who has the power and agency to incorporate this, does it take someone higher-up or can an individual engineer implement it?
    \item (if not the CEO) How would you pitch fairness to your CEO?
    \item Why does an RAI exist or not in the company?
    \item What would it take to hire someone to work only on RAI?
    \item What would it take to add RAI to a KPI or OKR or some kind of employee evaluation criteria?
    \item What would it take to incorporate RAI early into a product development lifecycle?
    \item What would it take to spend more resources on RAI?
    \item What do you think the company would be willing to compromise on for RAI, e.g., deployment delays, performance hits, etc.
    \item How has the prioritization of RAI changed over time, as the economy changes, as the company size changes?
    \item Can you tell me about a time where an RAI concern in a product occurred, and what factors made it feel significant enough to do something about?

\end{itemize}

\section{Actions for each Stakeholder}

\begin{figure*}
    \centering
    \includegraphics[width=0.9\textwidth]{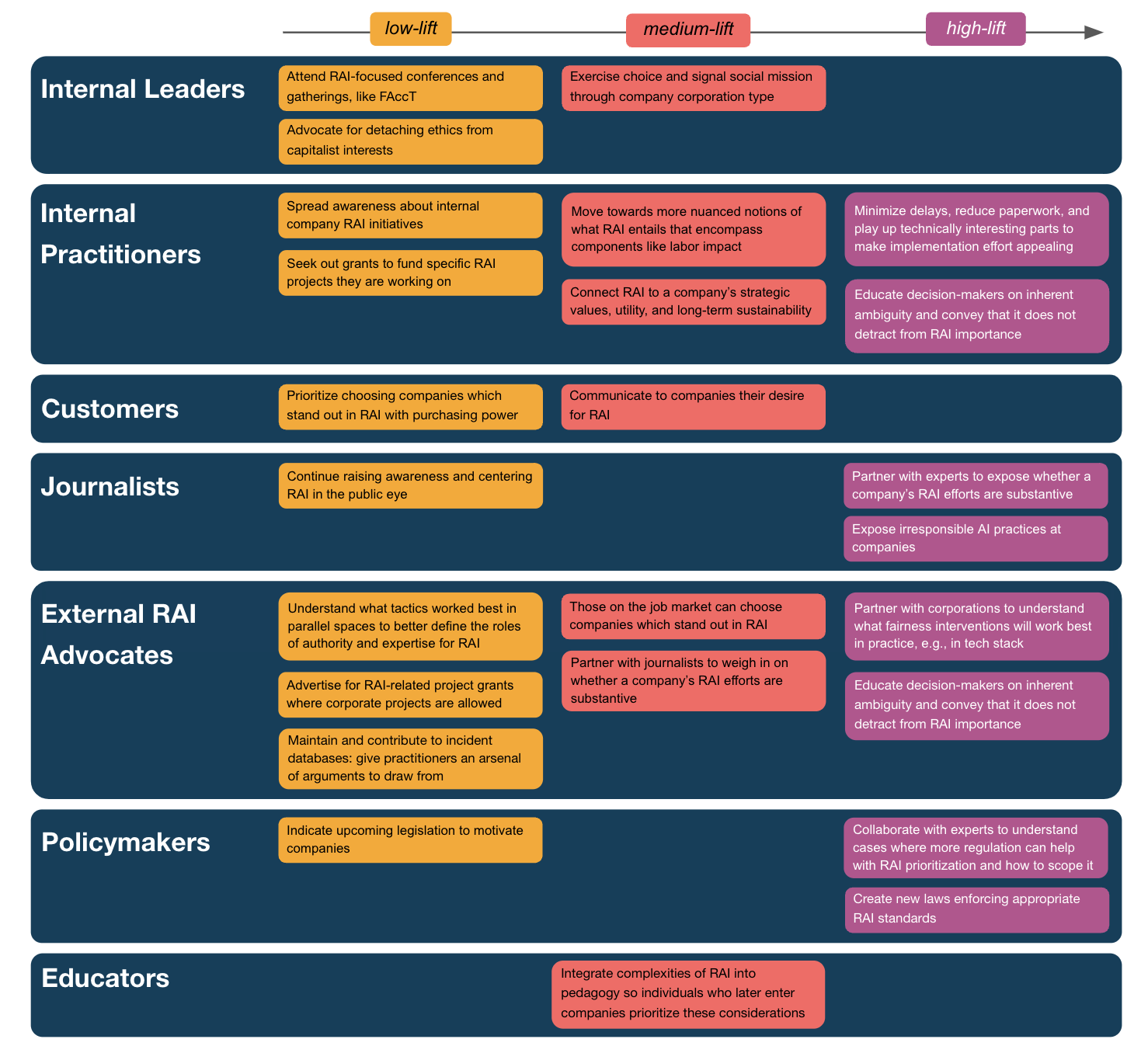}
    \caption{The strategies of action accessible to each key stakeholder group, organized according to level of effort.}
    \label{fig:stakeholder_summary}
\end{figure*}

In Fig.~\ref{fig:stakeholder_summary} we rearrange Fig.~\ref{fig:summmary} to showcase how different stakeholder groups can act on the strategies discussed in our work. These are only some of the strategies available for each key stakeholder group. We organize them loosely according to the level of effort required (low-lift, medium-lift, high-lift). It should be noted that the level of effort and level of effectiveness of these stakeholder action items is not vertically comparable. In other words, some stakeholders may have more power to affect change than others, and some action items marked as `low-lift' for certain groups may in fact require more effort than those labeled `high-lift' for other groups. We also caveat that the categorization of the amount of lift required for each item is itself very variable, i.e., the same action could vary in amount of lift depending on how it is executed on. We simply use this organizational scheme as a starting point for understanding the strategies that a diverse array of stakeholders can employ.

\end{document}